\documentclass[reprint,
    aps,
    superscriptaddress,
    10pt]{revtex4-2}
\usepackage{algorithm} 
\usepackage{algorithmic}
\usepackage{float}
\usepackage{graphicx}  % needed for figures
\usepackage{xcolor}
\usepackage{bm}        % for math
\usepackage{braket}
\usepackage{amssymb}   % for math
\usepackage{epstopdf}
\usepackage{xfrac} 
\usepackage{cancel}
\usepackage{soul}
\usepackage{amsmath}
\usepackage{amsthm}
\usepackage{amsfonts}
\usepackage{bbm}
\usepackage{comment}
\usepackage{dsfont}
\usepackage{physics}
\usepackage{setspace}
\usepackage[utf8]{inputenc}
\usepackage[english]{babel}
\usepackage[T1]{fontenc}
\usepackage{amsmath,amsthm,amsfonts,amssymb,xcolor}
\definecolor{darkblue}{rgb}{0.1,0.2,0.6}
\definecolor{darkred}{rgb}{0.8,0.1,0.2}
\definecolor{darkgreen}{rgb}{0.31,0.62,0.24}
\definecolor{bleudefrance}{rgb}{0.19, 0.55, 0.91}
\usepackage[colorlinks,citecolor=darkblue,linkcolor=darkblue,urlcolor=darkblue]{hyperref} 
\usepackage{enumerate}
\usepackage{setspace}
\usepackage{url}  % This makes \url work
\usepackage{mathrsfs}
\usepackage{mathtools} 
\usepackage{orcidlink}

\date{\today}

\renewcommand{\ket}[1]{\vert #1 \rangle}
\renewcommand{\bra}[1]{\langle #1 \vert}
\renewcommand{\mel}[3]{\langle #1 \vert #2 \vert #3 \rangle}
\newcommand{\ketnorm}[1]{\langle #1 \vert #1 \rangle}
\renewcommand{\braket}[2]{\langle #1 \vert #2 \rangle}

\DeclareMathOperator*{\argmin}{argmin}

\begin{document}
\title{A Biorthogonal Neural Network Approach to Two-Dimensional Non-Hermitian Systems}

\author{Massimo Solinas}
\affiliation{Institute for Theoretical Physics, ETH Zürich, 8093, Switzerland}
\author{Brandon Barton}
\affiliation{Institute for Theoretical Physics, ETH Zürich, 8093, Switzerland}
\author{Yuxuan Zhang}
\affiliation{Department of Physics and Centre for Quantum Information and Quantum Control, University of Toronto, 60 Saint George St., Toronto, Ontario M5S 1A7, Canada}
\affiliation{Vector Institute, W1140-108 College Street, Schwartz Reisman Innovation Campus Toronto, Ontario M5G 0C6, Canada}
\author{Jannes Nys}
\affiliation{Institute for Theoretical Physics, ETH Zürich, 8093, Switzerland}
\author{Juan Carrasquilla}
\affiliation{Institute for Theoretical Physics, ETH Zürich, 8093, Switzerland}

\begin{abstract}
Non-Hermitian quantum many-body systems exhibit a rich array of physical phenomena, including non-Hermitian skin effects and exceptional points, that remain largely inaccessible to existing numerical techniques. In this work, we investigate the application of variational Monte Carlo and neural network wavefunction representations to examine their ground-state (the eigenstate with the smallest real part of the energy) properties. Due to the breakdown of the Rayleigh-Ritz variational principle in non-Hermitian settings, we develop a self-consistent symmetric optimization framework based on variance minimization with a dynamically updated energy estimate. Our approach respects the biorthogonal structure of left and right eigenstates, and is further strengthened by exploiting system symmetries and pseudo-Hermiticity. 
Tested on a two-dimensional non-Hermitian transverse field Ising model endowed with a complex longitudinal field, our method achieves high accuracy across both parity-time symmetric and broken phases. Moreover, we propose novel optimization routines that address the challenges posed by exceptional points and provide reliable convergence to the ground state in regimes where standard variational techniques fail. Lastly, we show, through extensive numerical evidence, that our method offers a scalable and flexible computational tool to investigate non-Hermitian quantum many-body systems, beyond the reach of conventional numerical techniques such as the density-matrix renormalization group algorithm.
\end{abstract}

\maketitle

% \section{Introduction}
\paragraph*{\textbf{Introduction.}}
In quantum mechanics, the dynamics of a closed quantum system is fundamentally governed by a Hermitian Hamiltonian. However, in realistic settings where we have only limited access to a subsystem, non-Hermiticity (NH) naturally emerges. This occurs in many scenarios, such as photonic systems with losses~\cite{el2007theory}, open quantum systems subject to dissipation~\cite{gorini1976completely,lindblad1976generators}, or systems involving measurements and postselection~\cite{dalibard1992wave,dum1992monte,carmichael1993open}, where probability flow is no longer conserved. 
Although the formalism of NH quantum mechanics dates back to the 1950s~\cite{yang1952statistical}, the equilibrium and none-quilibrium properties of NH many-body systems remain a frontier of quantum physics. 
Intriguingly, the non-Hermicity in such systems leads to unconventional phase transitions and unique phenomena~\cite{fisher1978yang,hatano1996localization,hatano1997vortex,ashida2020non}, such as exceptional points~\cite{bender1969anharmonic,bergholtz2019exceptional}, the NH skin effect~\cite{yao2018edge,song2019non,lee2019anatomy,liang2022dynamic}, exotic supersonic modes in out-of-equilibrium systems~\cite{ashida2018full,dora2020quantum}, novel topological phases~\cite{rudner2009topological,lin2023topological,RevModPhys.93.015005}, and entanglement behavior unlike their Hermitian counterparts~\cite{gopalakrishnan2021entanglement,kawabata2023entanglement}.

Beyond theoretical advancements, NH quantum mechanics has seen numerous experimental demonstrations across diverse platforms, including photonic~\cite{guo2009observation,ruterObservationParityTime2010,regensburgerParityTimeSynthetic2012}, matter-light~\cite{PhysRevLett.117.123601,pengAntiparityTimeSymmetry2016}, and electronic systems~\cite{RevModPhys.93.015005,ashidaNonHermitianPhysics2020,pikulin2012topological,pikulin2013two,mi2014x}, highlighting its broad relevance and potential for technological innovation. Though early experiments primarily focused on small, few-particle systems~\cite{guo2009observation,ruter2010observation,feng2011nonreciprocal}, recent progress has started to reveal NH phenomena in strongly correlated settings~\cite{zhangObservationNonHermitianSupersonic2024a,PhysRevLett.124.250402}. This growing interface between non-Hermiticity and strong many-body interactions will continue to uncover even more exotic effects beyond those found in single-particle NH systems~\cite{yoshidaNonHermitianFractionalQuantum2019}.

Despite the growing interest in NH quantum many-body systems, theoretical and computational studies have been limited to non-interacting systems~\cite{RevModPhys.93.015005}, interacting models amenable to analytical techniques~\cite{RevModPhys.93.015005,fukui1998breakdown,PhysRevResearch.4.L012006,limaSpinTransportNonHermitian2023,PhysRevB.107.235153,medenMathcalPTsymmetricNonHermitianQuantum2023,buvca2020bethe,yamamotoTheoryNonHermitianFermionic2019}, or small systems accessible via exact diagonalization~\cite{RevModPhys.93.015005,yoshidaNonHermitianFractionalQuantum2019,makStaticsDynamicsNonHermitian2024,PhysRevB.106.L121102}. Studies of higher-dimensional or strongly correlated systems remain rare and are typically restricted to one-dimensional settings where density matrix renormalisation group (DMRG) and matrix product state (MPS) methods excel~\cite{PhysRevLett.97.110603,PhysRevResearch.4.L012006,zhongDensitymatrixRenormalizationGroup2024,Shen2024NonHermitian,guoVariationalMatrixProduct2022}. Meanwhile, two-dimensional correlated systems present a challenge for tensor-network and quantum Monte Carlo techniques.
For example, quantum Monte Carlo methods are hindered by the sign problem, which is expected in most NH systems~\cite{panSignProblemQuantum2024}, though exceptions exist~\cite{PhysRevB.104.125102,yu2024non}. As a result, strongly interacting, NH systems beyond one dimension remain largely unexplored~\cite{sayyadTransferLearningHermitian2024}, especially beyond the analytically tractable limit.

Nevertheless, variational Monte Carlo (VMC) combined with machine learning-based approaches, such as neural-network quantum states (NQS), have shown remarkable potential in overcoming some of these limitations for Hermitian systems, both in equilibrium~\cite{
carleo2017solving,hibat-allahRecurrentNeuralNetwork2020a,medvidovicNeuralnetworkQuantumStates2024,langeArchitecturesApplicationsReview2024,wuVariationalBenchmarksQuantum2023,chenEmpoweringDeepNeural2024,astrakhantsev2021broken} and non-equilibrium settings~\cite{PhysRevLett.125.100503,schmittQuantumPhaseTransition2022,PRXQuantum.4.040302, nys2024ab}. Also, VMC in combination with tensor-network methods has recently demonstrated to yield accurate ground-state solutions for quantum systems beyond one dimension~\cite{liu2025accurate, emonts2020variational, vieijra2021many}.
However, naively applying VMC to NH systems presents a fundamental obstacle: the conventional variational principle for ground-state energy minimization breaks down in the NH setting, due to a complex-valued energy spectrum and non-orthogonal eigenstates. To resolve this, we develop a self-consistent variance minimization framework tailored to VMC, along with a targeted initialization strategy that improves the performance. 

% \section{Variance minimization for non-Hermitian physics}
\paragraph*{\textbf{Variance minimization for non-Hermitian physics.}}

We adopt the biorthogonality formalism for NH quantum systems based on the biorthogonal structure of left and right eigenstates~\cite{Biorthogonal_QM, metric_framework_1}.

The right and left eigenstates $\ket{R_n}$ and $\ket{L_n}$ of a NH Hamiltonian $\hat{H}$ are defined by
\begin{equation}
    \hat{H}\ket{R_n} = \lambda_n\ket{R_n}, \quad \hat{H}^\dagger\ket{L_n} = \lambda_n^*\ket{L_n},
\end{equation}
with $\lambda_n \in \mathbb{C}$. Unlike Hermitian Hamiltonians, the eigenstates are not required to be orthogonal to each other and instead satisfy the biorthogonal relationship $\bra{L_n}R_m\rangle \propto \delta_{n,m}$. In this framework, a quantum state $\ket{\psi} \in \mathcal{H}$ is defined together with its dual complement $\ket{\tilde{\psi}}$, such that $\ket{\psi} = \sum_n c_n \ket{R_n} $ and $\ket{\tilde{\psi}} = \sum_n c_n \ket{L_n}$. The inner product is therefore modified so that the expectation value of an operator $\hat{O}$ is evaluated using $\mel{\tilde{\psi}}{\hat{O}}{\psi}/\braket{\tilde{\psi}}{\psi}$. 
\begin{figure*}[t]
    \centering
    \includegraphics[width=.95\textwidth]{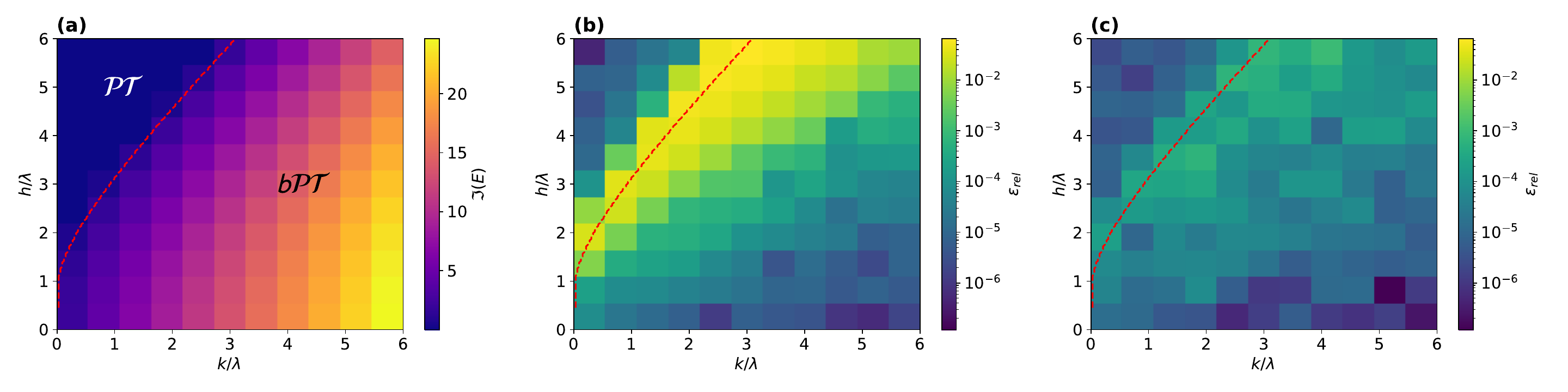}
    \caption{(a) Imaginary part of the ground-state energy of the NH-TFIM for a $N=9$ spin chain, computed using the symmetric self-consistent method together with the fixed start method. As the parameters vary, the system transitions from the $\mathcal{PT}$-symmetric phase to the $b\mathcal{PT}$ phase, passing through exceptional points, shown in red and obtained via exact diagonalization.
    (b, c) Relative energy error $\epsilon_{\text{rel}}$ between the exact ground state (from exact diagonalization) and the variational results from the two methods: energy as a parameter method, a previously introduced, (b) and self-consistent method that we develop in this work. (c). In both cases, the fixed start method is used to ensure convergence to the correct ground state. Our method exhibits improved convergence, highlighting that treating $\varepsilon$ as a free parameter can slow down the optimization by introducing saddle points into the loss landscape.}
    \label{fig:NH-TFIM}
\end{figure*}

Variational methods such as VMC rely on the Rayleigh-Ritz variational principle~\cite{ritz1909neue} to capture eigenstates. For NH systems, this principle is challenged: first, the notion of ordering eigenvalues breaks down for the complex eigenvalues; second, even when the energies of the system are all real, the non-orthogonality of the eigenstates forbids a proper formulation of the same variational principle. To overcome this challenge, we adopt the following variance-based loss function~\cite{siringo2005variational,umrigar2005energy}:
\begin{equation}\label{eq:loss}
    \mathcal{L}_R\left[\psi, \varepsilon\right] = \frac{\langle \psi \vert \hat{V}_R(\varepsilon) \vert \psi \rangle}{\langle \psi \vert \psi \rangle},
\end{equation}
where $|\psi \rangle$ is a (variational) quantum state, and $\hat{V}_R(E)$ is the right operator defined by
\begin{align}\label{Eq: var def}
    \hat{V}_R(\varepsilon) &= \left(\hat{H}^\dagger-\varepsilon^*\right)\left(\hat{H}-\varepsilon\right).
\end{align}
Here, $\varepsilon$ serves as an energy-like variable and its role will be discussed in detail below. Similarly, a left operator $\hat{V}_L(\varepsilon) = (\hat{H}-\varepsilon)(\hat{H}^\dagger-\varepsilon^*)$ can be defined, yielding a corresponding loss $\mathcal{L}_L[\tilde{\psi}, \varepsilon]$. These two operators $\hat{V}_{L,R}$ have the advantage of being Hermitian and positive semidefinite, so that their expectation value is real and bounded from below. 
Note that $\mathcal{L}_{R}$ reduces to the energy variance when $\varepsilon$ is the energy of the right state, i.e.\ $\varepsilon = \mel{\psi}{\hat{H}}{\psi}{\ketnorm{\psi}}$, and similar for $\mathcal{L}_{L}$. In this case, $\mathcal{L}_R$ ($\mathcal{L}_L$) vanishes whenever the right (left) state is an eigenstate of the Hamiltonian, even when both states have different energy or are not dual. However, crucial to our approach, we instead define $\varepsilon$ as the full biorthogonal expectation value, $\varepsilon = \mel{\tilde{\psi}}{\hat{H}}{\psi}/\braket{\tilde\psi}{\psi}$, in which case the two variances vanish when $\ket{\psi}$ and $\ket{\tilde{\psi}}$ are a pair of biorthogonal eigenstates.

While both $\hat{V}_{L,R}$ can be used as loss functions in VMC, two main challenges arise and must be addressed. First, while most prior studies treat $\varepsilon$ as an unconstrained variational parameter to update with gradient descent~\cite{two_steps_algo,zhangObservationNonHermitianSupersonic2024}, which we refer to as the energy-as-a-parameter method, this may introduce extra saddle points in the cost function landscape, hence hindering optimization (see Appendix~\ref{Appendix: e as parameter}). Second, since every eigenstate of the Hamiltonian satisfies the zero-variance condition, the number of local minima in the loss landscape is expected to grow exponentially with system size. However, our goal is to find the ground state, here defined as the the eigenstate with the smallest real part of the energy. 
In response to these two problems, we first introduce an efficient strategy to treat the variable $\varepsilon$, restoring its physical nature, and subsequently present two distinct methods for isolating the ground state in NH systems.

\paragraph{Self-consistent optimization.}
Our approach jointly optimizes the parameters of a quantum state and its dual complement by enforcing biorthogonality through $\varepsilon$, via the following loss function: 
\begin{equation}\label{Eq: loss function}
   \mathcal{L}\left[\psi, \tilde{\psi}\right] = \mathcal{L}_R\left[\psi, \varepsilon\right] +
    \mathcal{L}_L\left[\tilde{\psi}, \varepsilon\right],
\end{equation}
\begin{equation} \label{eq: epsilon}
    \varepsilon = \frac{\mel{\tilde{\psi}}{\hat{H}}{\psi}}{\braket{\tilde{\psi}}{\psi}}
\end{equation}

However, unlike the Hamiltonian in standard energy minimization, in this framework the operators $V_{L,R}$  depend on the trainable parameters of both $\ket{\psi}$ and $\ket{\tilde\psi}$ through $\varepsilon$, making gradient evaluation more computationally demanding and potentially destabilizing the optimization process. To address this, we propose to incorporate $\varepsilon$ self-consistently. Optimization proceeds by computing parameter updates of the two variational wavefunctions while keeping the shared energy estimate fixed during each update. This approach ensures that the wavefunctions converge to the correct left and right eigenstates of the NH Hamiltonian, with eigenvalues that remain complex conjugates, reflecting the expected spectral symmetry.

Although the algorithm yields a pair of biorthogonal states upon convergence in the non-degenerate case, subtleties can arise when eigenstates are degenerate.
In the latter scenario, the method may converge to a pair of degenerate states that, despite residing in the same subspace, do not manifest biorthogonality. Through biorthogonalization of the solution, one can nevertheless correctly evaluate the expectation value of an observable. 
However, for pseudo-Hermitian Hamiltonians, there exists an operator $\hat\eta$ such that $\ket{\tilde\psi} = \hat\eta \ket{\psi}$, which can be exploited to guarantee biorthogonality while optimizing only a single variational state.
The method is summarized in Algorithm~\ref{Algo: SC-opt}, and a more detailed explanation is provided in Appendix~\ref{app: sc algo}.

\paragraph{Targeting the ground state.}

To target the ground state with variance minimization, we adopt two complementary strategies, which we term as warm-start and fixed-start methods. The \emph{warm-start method} (also known as ``fine-tuning'' \cite{hernandes2025adiabatic, rende2024fine}),
inspired by the quantum adiabatic theorem, is effectively a transfer learning approach in ML and has already been employed in previous works~\cite{hibat2021variational, wu2019solving}. After decomposing the Hamiltonian into a Hermitian and NH part $\hat{H} = \hat{H}_h + k \hat{H}_{nh}$ with control parameter $k$, we start by optimizing for the ground state of the Hamiltonian with its NH part set to zero, $k=0$, using the standard variational principle. After, we gradually increase $k$ while every time performing a self-consistent optimization via Algorithm~\ref{Algo: SC-opt}. For each optimization at fixed $k$, we initialize the parameters with the ones obtained at the previous $k$. Similarly to the adiabatic theorem, we expect that if the parameter $k$ is varied sufficiently slowly and the spectral gap remains finite, the method will converge to the ground state of the non-Hermitian Hamiltonian~\cite{hibat2021variational,vzunkovivc2025variational}. 

In contrast, the \emph{fixed-start method} relies on a prior estimate of the ground state energy $E_0$, such as a lower bound of the energy spectrum, or an approximate mean-field solution. 

We use this initial estimate to fix $\varepsilon = E_0$ in the loss function, and optimize the variational wavefunction for $F$ steps. Subsequently, to avoid sudden shifts in the loss landscape the energy, $\varepsilon$ is smoothly transitioned to the self-consistent method over $T$ training steps via linear interpolation: 
\begin{equation}
    \varepsilon_{i+1} = \alpha_i E_0 + (1 - \alpha_i) \varepsilon_i, \quad \alpha_i = \frac{T - i}{T},
\end{equation}
with $i \in [1, T]$. Finally, after the transition period, the optimization continues self-consistently for another $M$ steps. Although the success of the fixed start method depends on the quality of the initial energy estimate, it requires only one optimization from any point in the phase diagram. The two methods can also be combined: the fixed-start method can provide an initial state for the warm-start method in some part of the NH phase diagram, so that the warm-start method can then explore the surroundings without starting from the Hermitian ground state.

\begin{figure*}[t]
    \centering
    \includegraphics[width=.92\textwidth]{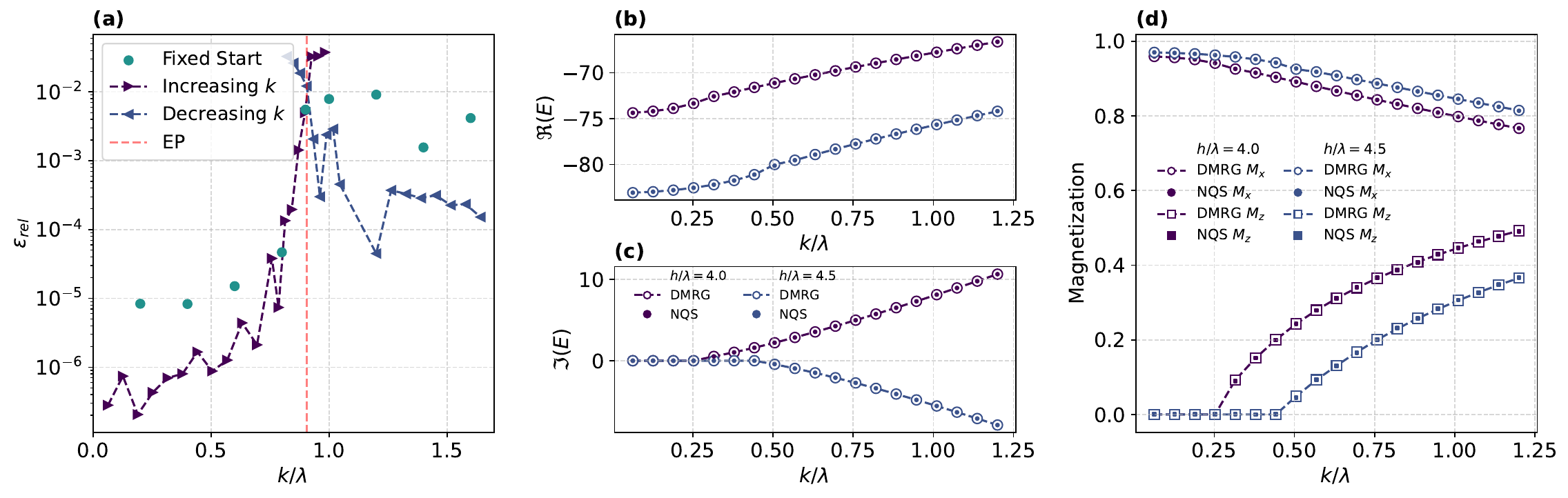}
    \caption{(a) Relative energy error obtained for a 1D spin chain of $N=20$ sites with a transverse field $h/\lambda = 3$. Each NQS is trained for $M = 5000$ steps. Both the forward (increasing $k$) and backward (decreasing $k$) sweeps were carried out using the self-consistent method combined with the warm start procedure. The backward sweep is initialized using the solution obtained from the fixed-start method. (b, c, d) Energy and magnetization for a 2D $N=6\times6$ lattice, using multiple NQS representations optimized with the self-consistent method. Each instance is trained for $M = 5000$ steps. The DMRG results are obtained using a maximum bond dimension of $\chi=1000$. In (b) and (c), the real and imaginary part of the ground state are shown as a function of the imaginary longitudinal field $k$. Panel (d) shows the magnetization along the $z$ and $x$ axis.}
    \label{fig: around EP and 6x6 warm}
\end{figure*}

\paragraph*{\textbf{Numerically capturing non-Hermitian phases.}}
Throughout this work, we consider the NH transverse-field Ising model (NH-TFIM):
\begin{equation} \label{Eq: NH-TFIM}
    \hat H = -\lambda \sum_{\langle i,j\rangle} \hat\sigma_i^z\hat\sigma_{j}^z - h \sum_i \hat\sigma_i^x - ik \sum_i\hat\sigma_i^z,
\end{equation}
where $h, k\in\mathbb{R}$, $\langle\cdot\rangle$ indicates the sum over nearest neighbors (see \cite{NH_TFIM_1, NH_TFIM_2, NH_TFIM_3, NH_TFIM_4, NH_TFIM_5} for different variations of the model). In the following analysis, we fix $\lambda=0.5$.
The NH-TFIM is a primary example exhibiting $\mathcal{PT}$ symmetry, where the parity operator is a global spin-flip as $\mathcal{\hat P} = \otimes_i \hat\sigma_i^x$, and the time reversal operator $\mathcal{\hat T}$ acts as complex conjugation. Moreover, the symmetry ensures pseudo-Hermiticity~\cite{PT_symm_QM}, such that the time-reversal operator provides the mapping from $\ket{\psi}$ to $\ket{\tilde\psi}$. The $\mathcal{PT}$-symmetry gives rise to two distinct phases of the system. In the unbroken $\mathcal{PT}$ phase, the eigenstates of $\hat{H}$ are also eigenstates of the $\mathcal{PT}$ operator, leading to a fully real energy spectrum and the system remains paramagnetic. Conversely, in the spontaneously broken $\mathcal{PT}$ phase ($b\mathcal{PT}$), where the eigenstates of $\hat{H}$ no longer coincide with those of the $\mathcal{PT}$ operator, the eigenvalues appear as complex-conjugate pairs and the system shows ferromagnetic order. The transition between these two phases occurs at an exceptional point (EP), 
and the imaginary transverse field shifts the position of the quantum phase transition along these~\cite{NH_TFIM_perturb, NH_TFIM_3}, defining a clear boundary between two distinct phases in the phase diagram.  Unlike the Hermitian counterpart, where phase transitions and non-analytical behavior occur in the thermodynamic limit, a sharp transition can occur in finite-size systems~\cite{weiUniversalCriticalBehaviours2017}. 
 This can be seen in Fig.~\ref{fig:NH-TFIM}(a). 

Throughout this work, we use a Restricted Boltzmann Machine (RBM) \cite{melko2019restricted} representation of the quantum state in the $\hat\sigma_z$ basis, and similarly for its biorthogonal counterpart. Moreover, to obtain a better convergence we use Stochastic Reconfiguration (SR) to precondition the gradients~\cite{beccaQuantumMonteCarlo2017}. We provide more detail about the used architecture and the use of SR for variance minimization in Appendices~\ref{app:rbm} and~\ref{Appendix: SR for variance}.

As a first test, we benchmark our self-consistent optimization approach against the previously proposed energy-as-a-parameter method (see Appendix~\ref{Appendix: e as parameter}). We find that our method accurately recovers the ground state across the entire phase diagram, in both the $\mathcal{PT}$-symmetric and broken-$\mathcal{PT}$ phases. Thanks to the improved cost function landscape, it achieves up to an order of magnitude improvement in accuracy compared to the previously used approach, as shown in Fig.~\ref{fig:NH-TFIM}(b,c).

Learning the ground state near EPs, at the paramagnet-to-ferromagnet quantum phase transition, is most challenging. This challenge stems from two key phenomena. First, the pathological nature of exceptional points leads to the coalescence of eigenstates, rendering them effectively indistinguishable. Second, the proximity to a quantum phase transition causes the energy gap between the ground and second excited states to shrink significantly (see Appendix~\ref{Appendix: gap}). Near this symmetry-breaking point, the optimization landscape develops competing minima associated with nearly degenerate energy levels. As a result, even small optimization inaccuracies can cause convergence to an incorrect state rather than the true ground state. We further explore this behavior in Appendix~\ref{Appendix: nh estimator}.

To address this challenge, we introduce a combined method that utilizes both the fixed and warm start, and thereby overcomes the individual limitations of the two methods. We approach exceptional points from two directions, starting from points within the $\mathcal{PT}$ and $b\mathcal{PT}$ phases. Starting in the $\mathcal{PT}$-phase towards the direction of the EP, we use the warm-start approach. Conversely, starting from the $b\mathcal{PT}$-phase, we first use the fixed-start to obtain a good initial state far away from the EP, and then proceed with the warm-start in the direction towards the EP by decreasing the control parameter on the NH term in the Hamiltonian, as shown in Fig.~\ref{fig: around EP and 6x6 warm}(a). This joint approach allows us to closely approach the quantum phase transition near the EP, where multiple states are nearly degenerate.

\paragraph*{\textbf{ Scaling two-dimensional systems.}}

\begin{figure*}[t]
    \centering
    \includegraphics[width=.92\textwidth]{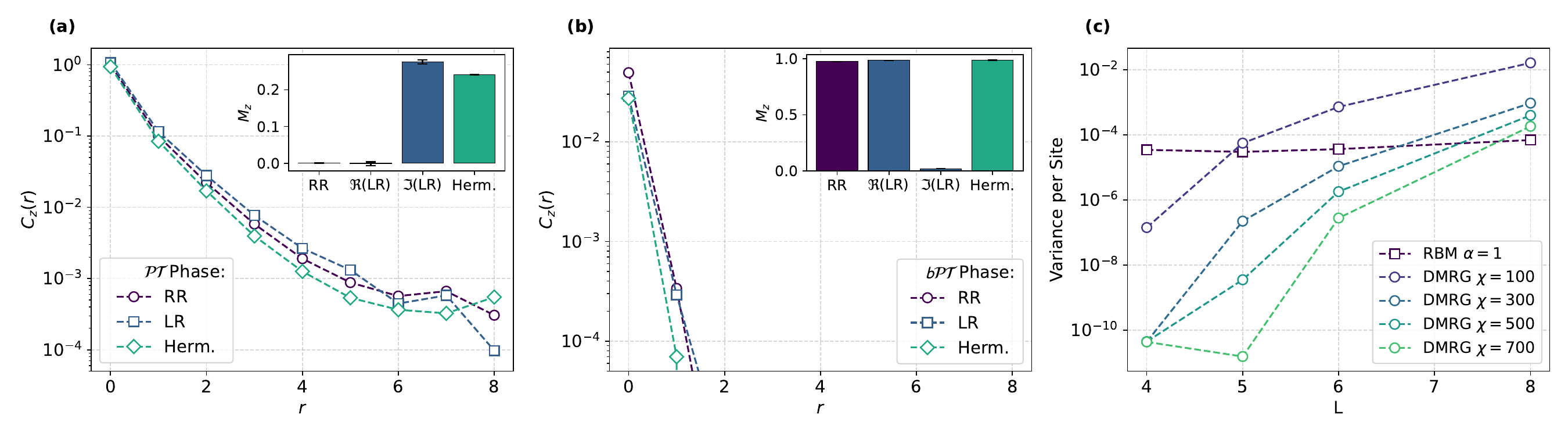}
    \caption{Panels (a, b) show the connected correlation function computed from the ground state of the NH-TFIM on a $8\times8$ lattice, in the $\mathcal{PT}$-symmetric ($k/\lambda=0.5$, $h/\lambda=5.5$) and broken $\mathcal{PT}$ phases ($k/\lambda=2$, $h/\lambda=1$), respectively. Expectation values are evaluated using both the standard RR and  biorthogonal LR formulation. For comparison, the correlations are also computed using the ground state of a Hermitian counterpart of the NH-TFIM, where the longitudinal field is made real via the substitution $k \to -i k$. The inset plots in panels (a, b) show the total magnetization on the $z$-axis for the Hermitian Hamiltonian and for the non-Hermitian one computed using both the RR and LR expectation values.
    In panel (b), correlations are truncated at $5 \cdot 10^{-5}$, where the Monte Carlo sampling error becomes comparable to the sampled values themselves. Panel (c) displays the variance per site (Eq.~\ref{Eq: var def}) of the right ground state, over different system sizes $N=L\times L$, obtained from both an NQS and a DMRG simulation with varying bond dimension $\chi$. For the NQS we adopt a RBM with parameter density $\alpha=1$.}
    \label{fig: CC and DMRG}
\end{figure*}

To fully demonstrate the advantage of NQS over traditional methods, we study the NH-TFIM on a 2D square lattice with periodic boundary conditions in both spatial directions, focusing on the paramagnet-to-broken-$\mathcal{PT}$ phase transition. To characterize $\mathcal{PT}$ symmetry breaking and the corresponding quantum phases, we compute the magnetization $ \hat{M}_\alpha = 1/N \sum_i \langle \hat\sigma_i^\alpha \rangle$ along the $\alpha=x,z$ axes, for a $6\times6$ system. As shown in Fig.~\ref{fig: around EP and 6x6 warm}, the NQS-based self-consistent approach yields results that closely match those obtained from DMRG across the entire parameter range for both the energies and the two magnetizations. Moreover, the behavior of the magnetization along the $z$-axis provides clear insight into the nature of the quantum phase transition occurring at the exceptional points. In the $\mathcal{PT}$-symmetric phase, the system exhibits a paramagnetic phase, characterized by a vanishing $M_z$-magnetization and a large $M_x$-magnetization. However, upon crossing the exceptional point, where the ground state energy becomes complex, the $M_z$-magnetization acquires a finite value, indicating the breaking of $\mathcal{PT}$ symmetry and a qualitative change in the nature of the eigenstates. This transition reveals that the longitudinal field effectively shifts the location of the quantum phase transition.

On the $N=8 \times 8$ lattice, we compute the ground state of the system in both the $\mathcal{PT}$-symmetric and broken $\mathcal{PT}$-symmetric phases, and characterize the connected correlations in these phases. We compute the connected spin–spin correlation function along the $z$ axis, defined as:
\begin{equation}\label{Eq: connected-correlation}
    C_z(r) = \frac{1}{N}\frac{1}{\mathcal{N}_r}\sum_i^N\left(\langle\hat\sigma^z_i\hat\sigma^z_{i+r}\rangle-\langle\hat\sigma^z_i\rangle\langle\hat\sigma^z_{i+r}\rangle\right),
\end{equation}
where $r$ is the Manhattan distance between two lattice sites, and $\mathcal{N}_r$ is the number of sites at that distance $r$ from a given site. We first evaluate the above expectation value using the biorthogonal inner product (LR). Second, we compute it using the standard quantum mechanical expectation value with only the right ground state (RR). Additionally, we performed the same calculation for a related Hermitian Hamiltonian, constructed by performing the transformation $k\to-ik$ on the NH parameter in the original model, thereby rendering the longitudinal field real \cite{vovrosh2021confinement}. Interestingly, in both the $\mathcal{PT}$-symmetric and the broken-$\mathcal{PT}$ phases (panels (a) and (b) of Fig.~\ref{fig: CC and DMRG}), the correlation functions computed from the ground states of the non-Hermitian and Hermitian models exhibit qualitatively similar behavior, decaying exponentially with distance. However, there is a qualitative difference in the physics of the two systems. In the Hermitian case, any finite longitudinal field results in a gapped energy spectrum~\cite{jalal2016topological}. 
In contrast, the non-Hermitian model further respects $\mathcal{PT}$ symmetry, which yields a spontaneous symmetry–breaking (SSB) transition at finite $k$, transitioning from a gapped phase to one with SSB-type correlations. 

This distinction becomes evident when examining the $z$-magnetization, as shown in the inset plots of Fig.~\ref{fig: CC and DMRG}~(a, b). In the Hermitian case, $M_z$ is nonzero at both points in the phase diagram as a result of the Hamiltonian's explicit breaking of the $\mathbb{Z}_2$ symmetry. Similarly, the broken $\mathbb{Z}_2$ symmetry in the NH-TFIM is observed in the finite value of $\abs{M_z}$ obtained with the LR estimator.

However, the non-Hermitian case shows richer physics, which can be observed in $\Re{M_z}$. While $\Re{M_z}$ is different from zero in the $b\mathcal{PT}$ phase, we observe that it vanishes in the $\mathcal{PT}$-symmetric phase (using both LR and RR estimators). This is a clear indication of the SBB of the $\mathcal{PT}$ symmetry. Indeed, since the non-Hermitian part of the Hamiltonian is proportional to $M_z$, the vanishing real part of $M_z$ is a direct consequence of the real energies in the $\mathcal{PT}$ phase.
We elaborate on LR expectation values in Appendix~\ref{Appendix: Biorthogonal Expectation Value}.

As a final benchmark, we show that the self-consistent method exhibits better variance scaling on large lattices compared to DMRG, as illustrated in Fig.~\ref{fig: CC and DMRG}(c). In fact, although DMRG achieves low variance for small system sizes, reflecting high accuracy, the variance per spin $\mathcal{L}/N$ increases significantly with the size of the system $N$, even when using a large bond dimension. In contrast, for our NQS method, the variance per spin remains nearly constant as the system size increases at a fixed number of trainable parameters per site. This improved scaling behavior indicates that the neural network-based approach becomes increasingly competitive for larger two-dimensional lattices.

\paragraph*{\textbf{Conclusions.}}
We have presented a VMC strategy that remains effective even when the Hamiltonian is non-Hermitian and the Rayleigh–Ritz energy principle no longer applies. By (i) reformulating the cost function as a variance within the biorthogonal framework, and (ii) updating the energy estimate self-consistently, the method yields reliable ground-state approximations using an NQS ansatz. Further, by carefully designing an initialization strategy, our framework outperforms previous methods, improving the energy accuracy by an order of magnitude. This advantage persists even in the vicinity of exceptional points. Benchmarks on one- and two-dimensional NH transverse-field Ising models show quantitative agreement with exact diagonalization and DMRG results. 

On two-dimensional lattices, our NQS representation exhibits improved scaling with system size, having the potential to outscale the widely-used MPS-based techniques such as DMRG~\cite{white1992density,white1993density,perez2006matrix,cirac2021matrix} or sequential circuits~\cite{haghshenas2021variational,zhang2022qubit,anand2023holographic,niu2022holographic,zhang2024sequential,chen2024sequential,chen2024sequential2}. These results establish neural-network quantum states, trained via self-consistent variance minimization, as a scalable tool for studying interacting, higher-dimensional non-Hermitian systems. Natural extensions include applying this approach to real-time dynamics as well as exploring different classes of systems, such as non-Hermitian fermionic models. Another promising direction is the adoption of more expressive ansätze, including autoregressive models or transformer-based architectures~\cite{hibat-allahRecurrentNeuralNetwork2020a,bennewitzNeuralErrorMitigation2022}, which may offer advantages in systems with extensive or long-range correlations.

\textit{During the finalization of this work, a related study appeared~\cite{wah2025many}, which explores neural-network representations of non-Hermitian Hamiltonians, focusing on benchmarking different network architectures with one-dimensional systems.}

\paragraph*{\textbf{Acknowledgments.}}
The authors would like to thank R. Chitra, M. Gonzalez and Y.B. Kim for helpful discussions. YZ was supported by the Natural Science and Engineering Research Council (NSERC) of Canada and acknowledges support from the Center for Quantum Materials and Centre for Quantum Information and Quantum Control at the University of Toronto. Resources used in preparing this research were provided, in part, by the Province of Ontario, the Government of Canada through CIFAR, and companies sponsoring the Vector Institute \url{www.vectorinstitute.ai/#partners}.

\bibliography{main}

\clearpage
\onecolumngrid
\appendix

\section{Neural-network architecture}\label{app:rbm}

In this work, we use a variational parametrization of the wavefunction
\begin{align}
    \vert \psi_{\theta} \rangle = \sum_{\bm{\sigma}} \psi_{\theta}(\bm{\sigma}) \vert \bm{\sigma} \rangle,
\end{align}
expanded on the basis of spin configurations $\bm{\sigma} = (\sigma_1, \sigma_2, \dots, \sigma_N)$, where $\sigma_i = \pm1$, and $N$ is the number of spins.
We employ a shallow neural network, called the restricted Boltzmann machine (RBM) \cite{melko2019restricted}, to parametrize the coefficients:
\begin{align}
    \psi_{\theta}(\boldsymbol{\sigma}) = e^{\boldsymbol{a}\cdot\boldsymbol{\sigma}} \times \prod_i 2\cosh\left[( W\boldsymbol{\sigma} + \boldsymbol{b})_i\right],
\end{align}
where $\bm{a}\in \mathbb{C}^N$ and $\bm{b}\in \mathbb{C}^{\alpha N}$ are visible and hidden biases respectively, $W \in \mathbb{C}^{\alpha N \times N}$ is a weight matrix, and $\alpha$ is a scaling factor of the input dimension. We take $\alpha=1$ throughout this work. The set $\theta = \{\bm{a},\bm{b}, W\}$ of variational parameters are optimized.

\begin{figure*}[t]
    \centering
    \includegraphics[width=1\textwidth]{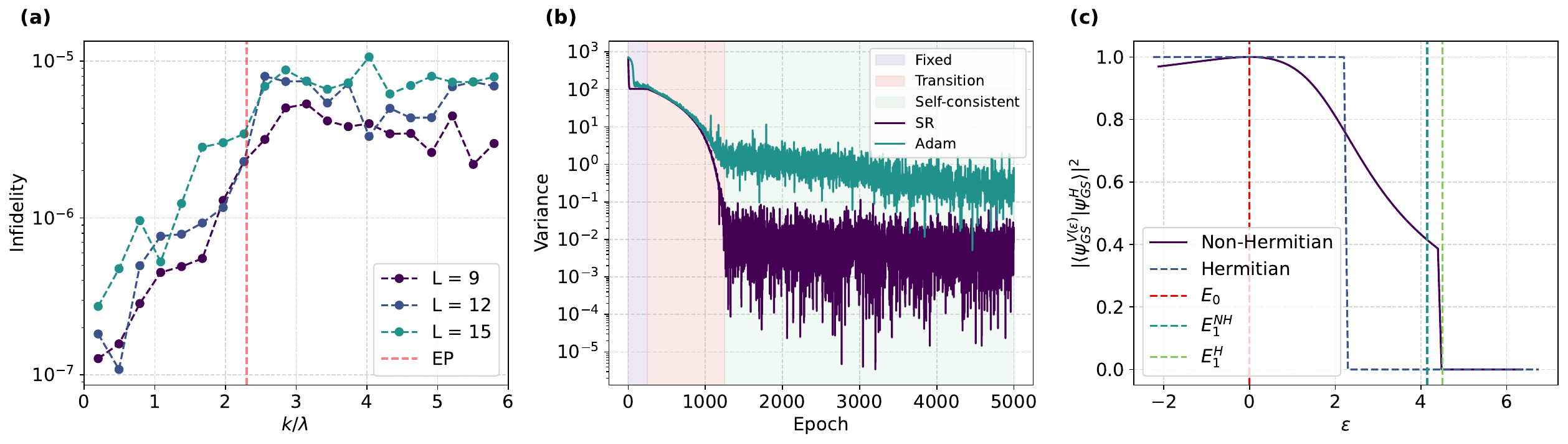}
    \caption{(a) Infidelity between the exact ground state and an RBM trained using infidelity optimization for different sizes of the chain $L$. The real transverse field was fixed to $h=2.5$. (b) Loss function $\mathcal{L}_R$ as a function of optimization steps using both SR and Adam optimizers. The three distinct optimization phases of the fixed-start method are indicated by different colors. (c) Fidelity between the ground state of the Hamiltonian $H$ and that of the variance operator $V(\varepsilon)$ as a function of the energy term $\varepsilon$. Results are shown for both the Hermitian and NH cases. The point $E_0$ denotes the ground state energy of each Hamiltonian, while $E_1^{NH}$ and $E_1^{H}$ correspond to the first excited states of the NH and Hermitian Hamiltonians, respectively.}
    \label{fig: opt_expressive_sr}
\end{figure*}

To determine whether the RBM can accurately represent ground states of the NH-TFIM across the phase diagram, we find optimal parameters for the RBM:
\begin{align}
    \theta^* &= \argmin_{\theta}\left[ \mathcal{I}(\phi, \psi_\theta) \right],\\
    \mathcal{I}(\phi, \psi_\theta) &= 1- \frac{|\langle \phi \vert \psi_\theta \rangle|^2}{\langle \phi \vert \phi \rangle \langle \psi_\theta \vert \psi_\theta \rangle},
\end{align}
where $\mathcal{I}(\phi, \psi_\theta)$ is the infidelity, $\psi_\theta$ is the RBM wavefunction, and $\phi$ is the ground state obtained from exact diagonalization on small system sizes. As illustrated in Fig.~\ref{fig: opt_expressive_sr}~(a), the RBM successfully captures the ground state in both phases, although its ability to represent ground states is slightly lower in the $b\mathcal{PT}$-phase. Furthermore, increasing the system size appears to have only a minor impact on infidelity. 

\section{Fixed-start method analysis}\label{app:warm_start}

We demonstrate the optimization procedure of the fixed start method in panel (b) of Fig.~\ref{fig: opt_expressive_sr}, where the loss function is shown in the three different regimes. The method starts with a first rough optimization of the parameters of the model during the fixed regime where $\varepsilon=E_0$ is kept fixed. During this period, the variance function rapidly converges. From here, the transition steps are crucial to move the minimum of the variance to the correct ground state, without changing the optimization landscape too rapidly.

In the case of the NH-TFIM we choose an initial guess for the energy the lower bound of the spectrum:
\begin{equation}
    E_0 = -N\left(\lambda\frac{\xi}{2}-h - ik\right),
\end{equation}
where $N$ is the number of spins and $\xi$ the number of nearest neighbors. 
This lower bound relies on the fact that the NH-TFIM in Eq.~\ref{Eq: NH-TFIM}, is composed of Pauli terms, which have eigenvalues $\pm1$.

\section{Stochastic Reconfiguration for Variance Minimization} \label{Appendix: SR for variance}

Stochastic Reconfiguration (SR) is a powerful natural gradient method that preconditions the gradient to accelerate convergence toward the ground state of a Hamiltonian. While SR has proven highly effective for energy minimization, a natural question that arises is how it might be adapted to improve convergence when minimizing the variance rather than the energy of the Hamiltonian. It is important to recognize that SR is rooted in imaginary-time evolution, a simple ground state search method based on applying the imaginary time propagator to an initial wave function: 
\begin{equation}
    \ket{E_0} \propto e^{-\tau\hat{H}}\ket{\psi}, \quad \text{for} ~~\tau\to+\infty,
\end{equation}
where $\ket{E_0}$ is the ground state of the system.

For this reason, understanding whether SR can be applied for variance optimization suggests considering whether the variance operator can be interpreted as an auxiliary Hamiltonian capable of generating an imaginary-time propagator. Whether this interpretation is valid depends crucially on two factors: whether the Hamiltonian is Hermitian, and the value of the energy term appearing in the variance operator. In the Hermitian case, the variance operator and the Hamiltonian share the same set of eigenvectors, as the variance operator effectively shifts the eigenvalues of the Hamiltonian by $-\varepsilon$ and then squares them. Consequently, as long as the energy term $\varepsilon$ is closer to the ground state energy $E_0$ than to any excited state $E_i$, the variance operator $V(\varepsilon)$ and the Hamiltonian $H$ will share the same ground state. In this regime, imaginary-time evolution can still drive the system toward the correct ground state. 

On the other hand, the situation is more intricate in the NH case. As shown in Fig.~\ref{fig: opt_expressive_sr}(c), due to the non-orthogonality of the eigenvectors of $H$, the variance operator and the Hamiltonian share a common eigenvector only when the energy term is set exactly equal to one of the system’s energies. Therefore, for imaginary-time evolution to be effective, the energy term must be dynamically updated at each iteration so that it progressively approaches the correct energy. In this context, the need for a self-consistent update becomes even more apparent. 

\section{NH Monte Carlo estimator} \label{Appendix: nh estimator}
NH quantum systems differ significantly from their Hermitian counterparts due to the absence of an orthogonal basis of eigenstates for the Hamiltonian. This fundamental distinction necessitates a redefinition of the quantum expectation value using a modified dual space. Consequently, it is essential to derive a suitable Monte Carlo estimator for the NH expectation value. In this case, two distinct wavefunctions are involved, denoted as: 
\begin{equation}
    \ket{\psi} = \sum_{\boldsymbol{\sigma}}\psi(\boldsymbol{\sigma})\ket{\boldsymbol{\sigma}}
\end{equation}
and its corresponding dual state
\begin{equation}
    \ket{\tilde\psi} = \sum_{\boldsymbol{\sigma}}\tilde\psi(\boldsymbol{\sigma})\ket{\boldsymbol{\sigma}},
\end{equation}
so that given an operator $\hat{O}$, its expectation value can be computed as:
\begin{equation}
    \langle\hat{O}\rangle= \frac{ \mel{\tilde{\psi}}{\hat{O}}{\psi} }{ \braket{\tilde{\psi}}{\psi} } 
    = \frac{ \sum_{\boldsymbol{\sigma}} \tilde{\psi}(\boldsymbol{\sigma})^* [\hat{O}\psi](\boldsymbol{\sigma}) }{ \sum_{\boldsymbol{\sigma}} \tilde{\psi}(\boldsymbol{\sigma})^* \psi(\boldsymbol{\sigma}) }
\end{equation}
To construct a proper Monte Carlo estimator, it is first needed to identify the probability distribution function $p(\boldsymbol{\sigma})$ from which the samples will be drawn. In the following, we propose two different options. Firstly, we can define:
\begin{equation}
    p(\boldsymbol{\sigma}) = \frac{ |{ \tilde{\psi}(\boldsymbol{\sigma}) \psi(\boldsymbol{\sigma}) |} }{ \sum_{\boldsymbol{\sigma}} |{ \tilde{\psi}(\boldsymbol{\sigma}) \psi(\boldsymbol{\sigma}) } |},
\end{equation}
such that
\begin{equation}\label{eq:mcestimator1}
    \langle{\hat{O}}\rangle = \frac{ \mathbb{E}_{\boldsymbol{\sigma} \sim p(\boldsymbol{\sigma})} \left[\frac{[\hat{O}\psi](\boldsymbol{\sigma})}{\psi(\boldsymbol{\sigma})} e^{i\phi(\boldsymbol{\sigma})} \right]}{ \mathbb{E}_{\boldsymbol{\sigma} \sim p(\boldsymbol{\sigma})} \left[ e^{i\phi(\boldsymbol{\sigma})}\right] } ,
\end{equation}
where
\begin{align}
    \phi = \textrm{arg}\left[\tilde{\psi}(\boldsymbol{\sigma})^* \psi(\boldsymbol{\sigma})\right].
\end{align}

Alternatively, one can sample from the probability distribution
\begin{equation}
    p(\boldsymbol{\sigma}) = \frac{ |{ \tilde{\psi}(\boldsymbol{\sigma}) }|^2 }{ \sum_{\boldsymbol{\sigma}} |{ \tilde{\psi}(\boldsymbol{\sigma}) }|^2 },
\end{equation}
such that
\begin{equation}
    \langle\hat{O} \rangle= \frac{ \mathbb{E}_{\boldsymbol{\sigma} \sim p(\boldsymbol{\sigma})} \left[\frac{[\hat{O}\psi](\boldsymbol{\sigma})}{\tilde{\psi}(\boldsymbol{\sigma})} \right]}{ \mathbb{E}_{\boldsymbol{\sigma} \sim p(\boldsymbol{\sigma})} \left[ \frac{\psi(\boldsymbol{\sigma})}{\tilde{\psi}(\boldsymbol{\sigma})} \right] }.
\end{equation}
A similar result can be obtained using the right state as probability distribution:
\begin{equation}
    p(\boldsymbol{\sigma}) = \frac{|{ \psi(\boldsymbol{\sigma}) }|^2 }{ \sum_{\boldsymbol{\sigma}} |{ \psi(\boldsymbol{\sigma}) }|^2 }.
\end{equation}

Both proposed non-Hermitian estimators satisfy the zero-variance principle, meaning that their variance vanishes when $\ket{\psi}$ is an eigenstate of $\hat H$.

For our ansatz we choose the two states to have the same functional form, but independent variational parameters
\begin{equation}
    \ket{\psi} \simeq \ket{\psi(\theta}), \quad \ket{\tilde\psi} \simeq \ket{\psi({\theta'})}.
\end{equation}

In the case of the NH-TFIM used in this work, we have $\ket{\tilde{\psi}} = \ket{\psi}^*$, so that the above expressions require only one set of parameters. However, when the two wavefunctions $\ket{\tilde{\psi}}$ and $\ket{\psi}$ are nearly orthogonal, the denominator in the estimator becomes small. In such cases, estimating a small quantity using Monte Carlo methods is challenging.

In regions of the phase diagram where the overlap between the right and left ground states is small, what we refer to as ``low-fidelity regions'', our self-consistent method performs slightly worse. To gain a better understanding of this behavior, we carry out a finite-size scaling analysis of the fidelity between these two states using exact diagonalization. As shown in Fig.~\ref{fig: fidelity_scaling_mag}(a,b), the overlap decreases exponentially with system size, with the steepest decline occurring at and near exceptional points within the $b\mathcal{PT}$ region. This suppression likely explains the reduced accuracy of our method in these areas, as the fidelity estimator in the denominator requires a significantly larger number of samples to achieve the same level of precision.

\section{Biorthogonal Expectation Value} \label{Appendix: Biorthogonal Expectation Value}

\begin{figure*}[t]
    \centering
    \includegraphics[width=1\textwidth]{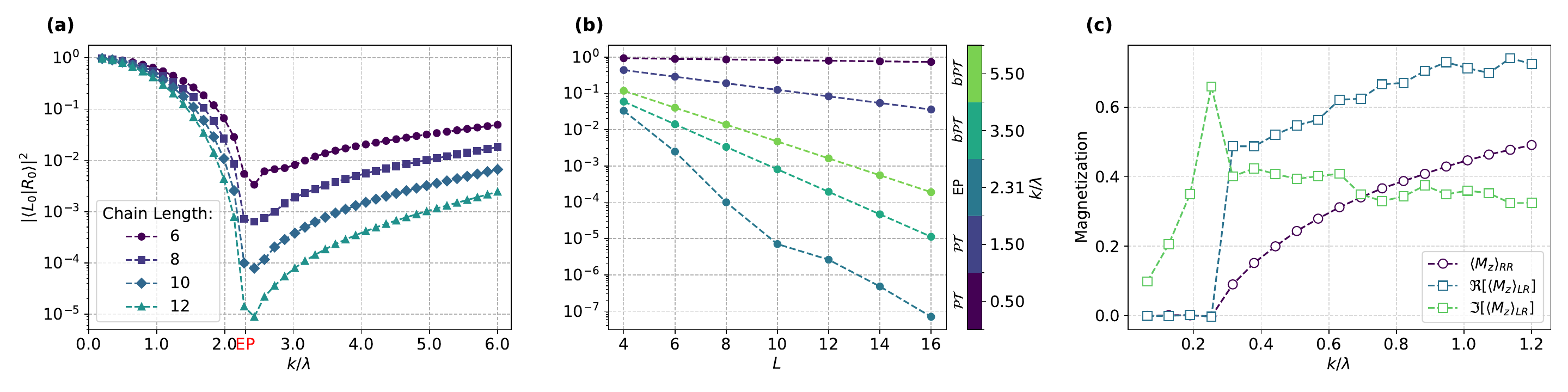}
    \caption{(a, b) Scaling of the fidelity between the normalized left and right ground states as a function of the chain length $L$ (a) and the imaginary field $k$ (b). The scaling was obtained with exact diagonalization, fixing $h/\lambda = 5$. (c) Magnetization along the $z$ axis, computed using both the standard expectation value (RR), which involves only the right ground state, and the biorthogonal expectation value (LR), which involves both the left and right ground states. Due to the NH nature of the Hamiltonian, the expectation values are generally complex; we therefore show the real and imaginary parts separately. All results are obtained for the ground state of the NH-TFIM on a $6 \times 6$ lattice, using the symmetric self-consistent method.}
    \label{fig: fidelity_scaling_mag}
\end{figure*}

As previously discussed, biorthogonal quantum mechanics modifies the definition of the inner product in Hilbert space to enable a consistent probabilistic interpretation. In standard quantum mechanics, the inner product ensures that the expectation value of a Hermitian operator is real. However, this guarantee no longer holds when expectation values are computed using both $\ket{\psi}$ and $\ket{\tilde\psi}$.

While we find good agreement between the standard (RR) and biorthogonal (LR) expectation values for the connected correlation function (see panels (a, b) of Fig.\ref{fig: CC and DMRG}), the behavior of the magnetization is markedly different, as shown in Fig.\ref{fig: fidelity_scaling_mag}(c). Notably, the magnetization becomes complex even within the $\mathcal{PT}$-symmetric phase, where the ground state remains stationary. Moreover, while the RR expectation value yields a continuously varying order parameter, the real part of the LR expectation exhibits a sudden jump at the exceptional point, suggesting the presence of a first-order phase transition.

The question of how to properly define and interpret observables in NH quantum mechanics remains an open and actively debated topic \cite{sim2025observables}. Some proposals argue that only operators with real expectation values should qualify as observables within the biorthogonal framework \cite{brody2016consistency}.

\section{Spectral Gap in the NH-TFIM}\label{Appendix: gap}
 To gain a deeper understanding of the quantum phase transition happening at the EPs, we investigate the energy gap $\Delta$ as a function of the system size through exact diagonalization. However, since the energies are complex in the $b\mathcal{PT}$-phase and both the ground state and the first excited state coalesce at the EP, the gap must be redefined in a manner that accounts for these two properties. Specifically, we define the gap as the modulus of the difference between the ground state energy and the closest energy, excluding the one of the first excited state. That is:
\begin{equation}\label{Eq: spectral gap}
\Delta = \min_{E \in \mathcal{E}} |E - E_0|,
\end{equation}
where $E_0$ is the ground state energy, and $\mathcal{E} = \{ E_n \}_{n \geq 2}$ is the set of energy levels excluding the first excited state. 

\begin{figure}
    \centering
    \includegraphics[width=1\linewidth]{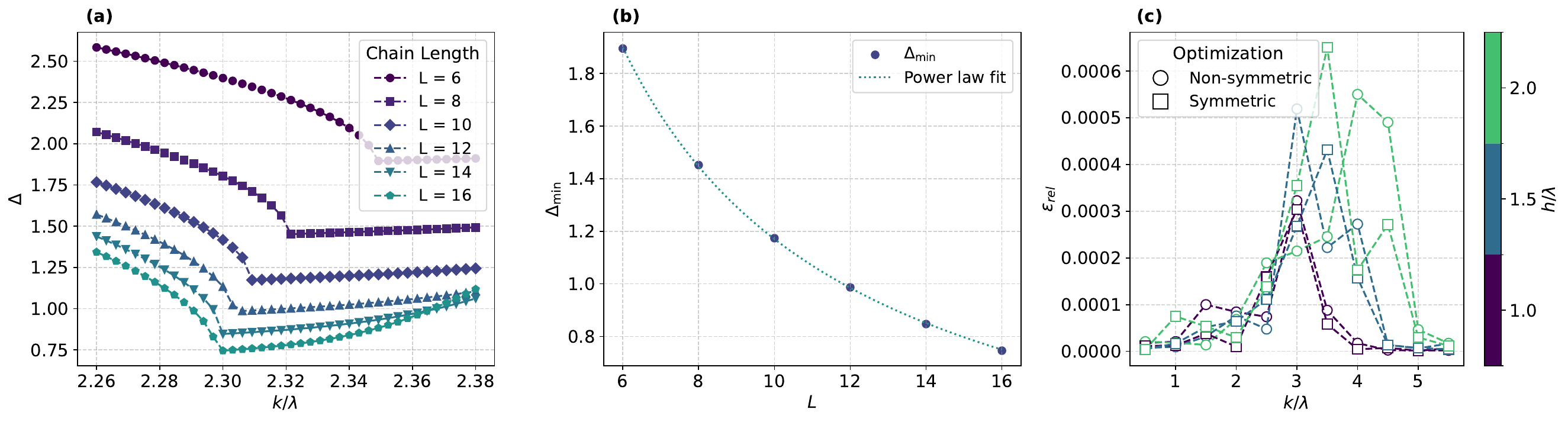}
    \caption{(a) The spectral gap as defined in Eq.~\ref{Eq: spectral gap} as a function of $k/\lambda$. (b) The minimum of the gap $\Delta_{min}$ for each system size $L$ fitted with a power law function. The scaling was obtained through exact diagonalization of the 1D NH-TFIM with the real field set to $h = 2.5$. (c) Relative error between the ground state obtained via exact diagonalization and the one obtained through self-consistent optimization, for a one-dimensional system of 9 spins. The non-symmetric optimization is performed by independently training two wave functions, $\ket{\psi}$ and $\ket{\tilde\psi}$, while the symmetric scheme trains only $\ket{\psi}$ and exploits the $\mathcal{PT}$ symmetry to construct $\ket{\tilde\psi}$ from it.}
    \label{fig: gap study}
\end{figure}

As shown in Fig.~\ref{fig: gap study}~(a, b), our results indicate that as the system size increases, the gap closes at the EP following a power-law behavior with an exponent $\alpha=-0.948$. This suggests that in the thermodynamic limit ($L\to+\infty$), the gap approaches exactly zero, signaling the presence of a quantum phase transition. 

Therefore, at the EP, we observe two distinct effects: first, the ground state and the first excited state become the same due to the spontaneous $\mathcal{PT}$-symmetry breaking and the NH nature of the Hamiltonian, and second, the ground state itself becomes degenerate due to the quantum phase transition.

\section{Self-consistent algorithm} \label{app: sc algo}

The general procedure used to optimize the two states $\ket{\psi}$ and $\ket{\tilde\psi}$ is summarized in Algorithm~\ref{Algo: SC-opt}. Below, we provide further motivation for why this procedure enforces the biorthogonality condition between the two states upon convergence.

The method begins by initializing two wavefunctions, $\ket{\psi}$ and $\ket{\tilde\psi}$, such that they are not biorthogonal to one another initially. Since the operators $\hat{V}_{R/L}(\varepsilon)$ are Hermitian and positive semi-definite, their spectra are real and non-negative. By minimizing the expectation value of $\hat{V}_R(\varepsilon)$ with respect to the trainable parameters of $\ket{\psi}$, and $\hat{V}_L(\varepsilon)$ with respect to ones of $\ket{\tilde\psi}$, the variational principle ensures that, upon convergence, the wavefunctions correspond to the ground states of their respective operators.

Furthermore, these wavefunctions are those whose energies are closest to $\varepsilon$ and $\varepsilon^*$, respectively, as the procedure minimizes the residuals: 
\begin{equation}
     \left\lVert(\hat H - \varepsilon)\ket{\psi}\right\lVert^2, \quad \left\lVert(\hat H^\dagger - \varepsilon^*)\ket{\tilde\psi}\right\lVert^2.
\end{equation}

If these residuals remain non-zero, it implies that the states are not exact eigenstates of $\hat{H}$ and $\hat{H}^\dagger$, and hence are not biorthogonal. To address this, a self-consistent update of the parameter $\varepsilon$ is performed using Eq.~\ref{eq: epsilon}, which iteratively moves $\varepsilon$ closer to the actual eigenvalue of the Hamiltonian. This process is repeated until the two wavefunctions converge to eigenstates of $\hat{V}_R$ and $\hat{V}_L$ that are also eigenstates of $\hat{H}$ and $\hat{H}^\dagger$, respectively. Importantly, since $\varepsilon$ is a shared quantity between the two loss functions, the two states converge to eigenstates of the Hamiltonian with energies that are complex conjugate pairs.

In practice, it is beneficial to update the value of $\varepsilon$ at every iteration to accelerate convergence. Once $\ket{\psi}$ and $\ket{\tilde\psi}$ become eigenstates of $\hat{H}$ and $\hat{H}^\dagger$ with eigenvalues $\varepsilon$ and $\varepsilon^*$, respectively, and if the corresponding eigenvalue is non-degenerate, then the states are automatically biorthogonal by definition. However, if the eigenvalue is degenerate, a more subtle situation arises. Suppose the states converge to
\begin{equation}
    \ket{\psi}=\left(a_1 \ket{R} + a_2\ket{R'}\right), \quad \ket{\tilde\psi} = \left(b_1\ket{L}+b_2\ket{L'}\right),
\end{equation}

where $\ket{R}$ and $\ket{R'}$ are eigenstates of $\hat{H}$ with eigenvalue $E$, $\ket{L}$ and $\ket{L'}$ are eigenstates of $\hat{H}^\dagger$ with eigenvalue $E^*$. While these states still yield the correct eigenvalue through Eq.~\ref{eq: epsilon}, they are not biorthogonal if $a_i\neq b_i$ for $i=1,2$. Consequently, computing the expectation value of another observable $\hat{O}$ that does not share the same eigenstates may result in incorrect biorthogonal predictions. To restore biorthogonality in this degenerate case, one must explicitly construct a biorthogonal pair of states by projecting onto the appropriate subspace and orthonormalizing the left and right eigenstates. 

On the other hand, if the Hamiltonian is pseudo-Hermitian and the corresponding metric operator $\hat{\eta}$ is known, the issue can be circumvented by optimizing only a single state $\ket{\psi}$ and constructing its dual using $\hat{\eta}$. Indeed, using the relation $\hat{\eta} \ket{R_n} = \ket{L_n}$, one can write, for any state $\ket{\psi}$:

\begin{equation}
\hat{\eta} \ket{\psi} = \sum_n c_n \hat{\eta} \ket{R_n} = \sum_n c_n \ket{L_n} = \ket{\tilde{\psi}}.
\end{equation}

If $\hat{\eta}$ is anti-linear, as is the case for the NH-TFIM, the result still holds. In this case, the coefficients can be written as $c_n = |c_n| e^{i\phi_n}$, yielding:

\begin{equation}
\hat{\eta} \ket{\psi} = \sum_n |c_n| e^{-i\phi_n} \hat{\eta} \ket{R_n} = \sum_n |c_n| e^{-i\phi_n} \ket{L_n} = \sum_n c_n e^{-2i\phi_n} \ket{L_n}.
\end{equation}
One can then exploit the gauge freedom in the definition of the left eigenstates by redefining:

\begin{equation}
\ket{L_n} \to e^{-2i\phi_n} \ket{L_n},
\end{equation}
such that $\ket{\tilde{\psi}} = \hat{\eta} \ket{\psi}$ remains valid even in the anti-linear case.

In this work, we exploit the $\mathcal{PT}$ symmetry of the Hamiltonian by optimizing only a single wavefunction. Nevertheless, we find that optimizing both states using the general method yields the same level of accuracy in the case of the NH-TFIM, where the ground state is non-degenerate, as shown in panel (c) of Fig.~\ref{fig: gap study}. This confirms the validity and robustness of the general self-consistent optimization scheme.

\begin{algorithm}[H] 
    \caption{Self-Consistent Optimization.}
    \label{Algo: SC-opt}
    \begin{algorithmic}
        \STATE Initialize two parametrized wavefunctions: $\ket{\psi_\theta}$ and $\ket{\tilde{\psi}_{\theta'}}$, and call $\Theta=\left(\theta,\theta'\right)$.
        \FOR{$m = 1$ to $M$}
            \STATE Update the energy with: 
            $\varepsilon = \frac{\bra{\tilde{\psi}_{\theta'}} \hat H \ket{\psi_\theta}}{\langle \tilde{\psi}_{\theta'}|\psi_\theta\rangle}.$
            \STATE Update the parameters with:
            $
            \Theta \gets \Theta - \lambda\nabla_{\Theta}\mathcal{L}[\psi_\theta, \tilde{\psi}_{\theta'}, \varepsilon]
            $
        \ENDFOR
    \end{algorithmic}
\end{algorithm}

\section{Energy as a parameter method} \label{Appendix: e as parameter}

In the energy as a parameter method, a variation on methods introduced from previous work using variational quantum algorithms \cite{two_steps_algo}, the energy term $\varepsilon$, appearing in the variance operator in Eq.~\ref{Eq: var def}, is updated iteratively during the optimization process using a simple gradient descent rule:
\begin{equation} \label{Eq: energy update par 1}
    \varepsilon_{new} = \varepsilon_{previous} - \lambda\frac{\partial\mathcal{L}_i(\theta_i, \varepsilon)}{\partial \varepsilon},
\end{equation}
in which $i$ can be either $L$ or $R$, and $\lambda$ is the learning rate. The optimization procedure is outlined in Algorithm~\ref{Algo: Parametrized-E Optimization}.

\begin{algorithm}[H] 
\caption{Parametrized-E Optimization}
\label{Algo: Parametrized-E Optimization}
\begin{algorithmic}
    \STATE Initialize one parametrized wavefunction $\ket{\psi(\theta_i)}$, with $i$ being either $L$ or $R$ 
    \FOR{$m = 1$ to $M$}
        \STATE Update the energy with $\varepsilon \gets \varepsilon - \lambda\frac{\partial\mathcal{L}_i}{\partial \varepsilon}$
        \STATE Update the parameters with:
        $\theta_i \gets \theta_i - \lambda\frac{\partial\mathcal{L}_i(\varepsilon)}{\partial\theta_i}$
    \ENDFOR
\end{algorithmic}
\end{algorithm}

This method was originally introduced in \cite{two_steps_algo} with a slightly different implementation. In that work, the authors propose to update the real and imaginary parts of the energy parameter separately using a two-step procedure. In the first step, only the imaginary part of the energy is updated, while the real part remains fixed to a predetermined value, typically set lower than the expected ground-state energy. In the second step, both components of the energy are optimized via gradient descent. The primary motivation behind this method is to facilitate the convergence toward the ground state of the system. However, due to the exponential scaling of the number of energy levels with system size, this approach is not expected to remain effective when dealing with a large number of degrees of freedom, so that a warm or fixed start approach becomes necessary. 

Compared to our self-consistent optimization method, treating $\varepsilon$ as a trainable parameter can introduce extra saddle points in the optimization landscape, even when the Hamiltonian is Hermitian and $\varepsilon$ real. To illustrate this, consider a simple single-qubit Hamiltonian $H=\sigma_z$ and the following parametrization of the wave function:
\begin{equation}
    \ket{\psi(\theta)} = \sin\theta\ket{0}+\cos\theta\ket{1}.
\end{equation}
In this case, the cost function becomes:
\begin{equation}
    \mathcal{L}_R\left[\psi, \varepsilon\right] = 1+2\varepsilon\cos2\theta +\varepsilon^2.
\end{equation}

 For $\theta\in [0,\pi/2]$, it's easy to check that the global minimum is located at $\varepsilon = -1, \theta = \pi/2$. However, due to the presence of $\varepsilon$ as a free parameter, the cost function gives another saddle point at $\varepsilon = 0, \theta = \pi/4$, where the Hessian determinant becomes negative. In practice, the existence of these saddle points can hinder optimization by introducing plateaus and directions of negative curvature that confuse gradient-based update rules. Our self-consistent update method, on the other hand, avoids this problem and improves performance.
 
\end{document}